\documentclass[aps,prl,twocolumn,preprintnumbers,superscriptaddress]{revtex4}
\usepackage{amssymb}
\usepackage{bm}
\usepackage{amsmath}
\usepackage[dvips]{graphicx}
\usepackage{color}

\begin{document}

\title{Anisotropies in insulating La$_{2-x}$Sr$_x$CuO$_4$: angular resolved
photoemission and optical absorption}
\author{O. P. Sushkov}
\affiliation{School of Physics, University of New South Wales, Sydney 2052, Australia}
\author{Wenhui Xie}
\affiliation{Max-Planck-Institut fur Festkorperforschung, Heisenbergstrasse 1, D-70569
Stuttgart, Germany}
\author{O. Jepsen}
\affiliation{Max-Planck-Institut fur Festkorperforschung, Heisenbergstrasse 1, D-70569
Stuttgart, Germany}
\author{O. K. Andersen}
\affiliation{Max-Planck-Institut fur Festkorperforschung, Heisenbergstrasse 1, D-70569
Stuttgart, Germany}
\author{G. A. Sawatzky}
\affiliation{Department of Physics and Astronomy, University of British Columbia,
Vancouver B. C. V6T-1Z1, Canada}

\begin{abstract}
Due to the orthorhombic distortion of the lattice, the electronic hopping
integrals along the $a$ and $b$ diagonals, the orthorhombic directions, are
slightly different. We calculate their difference in the LDA and find $%
t_{a}^{\prime }-t_{b}^{\prime }\approx 8\,$meV. We argue that electron
correlations in the insulating phase of La$_{2-x}$Sr$_{x}$CuO$_{4}$, i. e.
at doping $x\leq 0.055,$ dramatically enhance the $\left( t_{a}^{\prime
}-t_{b}^{\prime }\right) $-splitting between the $a$- and $b$-hole valleys.
In particular, we predict that the intensity of both angle-resolved
photoemission and of optical absorption is very different for the $a$ and $b$
nodal points.
\end{abstract}

\maketitle


\section{Introduction}
The magnetic state of La$_{2-x}$Sr$_{x}$CuO$_{4}$ (LSCO) changes
tremendously with Sr doping. The three-dimensional antiferromagnetic 
N\'{e}el order identified~\cite{Keimer92} below $325\ \text{K}$ in the parent
compound disappears at doping $x\approx 0.02$ and gives way to the so-called
spin-glass phase which extends up to $x\approx 0.055$. In both, the N\'{e}el
and the spin-glass phase, the system essentially behaves as an Anderson
insulator and exhibits only hopping conductivity~\cite{Keimer92,ando02}.
Superconductivity then sets in for doping $x\gtrsim 0.055$, see Ref.~%
\onlinecite{Keimer92}. One of the most intriguing properties of LSCO is the
static incommensurate magnetic order observed at low temperature in elastic
neutron scattering experiments. This order manifests itself as a scattering
peak shifted with respect to the antiferromagnetic position. Very
importantly, the incommensurate order is a generic feature of LSCO.
According to experiments in the N\'{e}el phase, the incommensurability is
almost doping independent and directed along the orthorhombic $b$ axis~\cite%
{matsuda02}. In the spin-glass phase, the shift is also directed along the $b
$ axis, but scales linearly with doping~\cite{wakimoto99,matsuda00,fujita02}%
. Finally, in the underdoped superconducting region ($0.055\lesssim
x\lesssim 0.12$), the shift still scales linearly with doping, but it is
directed along the crystal axes of the tetragonal lattice~\cite{yamada98}.
In the present work we discuss only the insulating phase, $x\leq 0.055$. It
is clear that pinning of the diagonal spin structure is due to the
orthorhombic distortion of the crystal. 
A mechanism for pinning of the diagonal spin structure to the orthorhombic
$b$ axis was suggested in Refs.~\cite{sushkov05,luscher06,luscher07}. 
The mechanism has four components:\newline
1) Due to strong antiferromagnetic correlations, the minima of dispersion of
a mobile hole are at points $(\pm \pi /2,\pm \pi /2)$ of the Brillouin zone,
so the system can, to some extent, be considered as a two valley semiconductor.
\newline
2) At low temperature, each hole is trapped in a hydrogen-like bound state
near the corresponding Sr ion, the binding energy is about $10\,\text{meV}$
and the radius of the bound state is about 10 \AA .\newline
3) Due to the orthorhombic distortion, the diagonal hopping matrix elements $%
t_{a}^{\prime }$ and $t_{b}^{\prime }$ are slightly different, and this
makes the $b$ valley, $(-\pi /2,\pi /2)$, deeper than the $a$
valley, $(\pi /2,\pi /2)$. So all the hydrogen-like bound states are built
with holes from the $b$-valley.\newline
4) Each hydrogen like bound state creates a spiral distortion of the spin
background and the distortion is observed in neutron scattering.
So the state at $0.02 < x < 0.055$ is not a spin glass, it is a
disordered spin spiral.

In the present paper we calculate accurately the diagonal hopping matrix
elements and show that the difference is $t_{a}^{\prime }-t_{b}^{\prime
}\approx 8\,$meV. We also discuss the implications of the described physics
for the $p^{5}d^{9}\rightarrow p^{6}d^{8}$ optical absorption and for
angle-resolved photoemission (ARPES). The optical absorption probes the
charge distribution, and we predict that the absorption vanishes for
polarization along the orthorhombic $a$-direction. ARPES probes the spin
spiral, and we predict very different spectra in $a$ and $b$ nodal points.

\section{Anisotropy of $t^{\prime }$ due to orthorhombic distortion and
tilting}

The 2D $t-J$ model was suggested two decades ago to describe the essential
low-energy physics of high-$T_{c}$ cuprates~\cite{PWA,Em,ZR}. In its
extended version, this model includes additional hopping matrix elements $%
t^{\prime }$ and $t^{\prime \prime }$ to respectively 2nd and 3rd-nearest Cu
neighbors. The Hamiltonian of the $t-t^{\prime }-t^{\prime \prime }-J$ model
on the square Cu lattice has the form:%
\begin{eqnarray}
H &=&-t\sum_{\langle ij\rangle \sigma }c_{i\sigma }^{\dag }c_{j\sigma
}+t^{\prime }\sum_{\langle ij^{\prime }\rangle \sigma }c_{i\sigma }^{\dag
}c_{j^{\prime }\sigma }-t^{\prime \prime }\sum_{\langle ij^{\prime \prime
}\rangle \sigma }c_{i\sigma }^{\dag }c_{j^{\prime \prime }\sigma }  \notag \\
&+&J\sum_{\langle ij\rangle \sigma }\left( \mathbf{S}_{i}\mathbf{S}_{j}-{%
\frac{1}{4}}n_{i}n_{j}\right) .  \label{H}
\end{eqnarray}%
Here, $c_{i\sigma }^{\dag }$ is the creation operator for an electron with
spin $\sigma $ $(\sigma =\uparrow ,\downarrow )$ at site $i$ of the square
lattice, $\langle ij\rangle $ indicates 1st-, $\langle ij^{\prime }\rangle $
2nd-, and $\langle ij^{\prime \prime }\rangle $ 3rd-nearest neighbor sites. 
The spin operator is $\mathbf{S}_{i}={\frac{1}{2}}c_{i\alpha }^{\dag }%
\mathbf{\sigma }_{\alpha \beta }c_{i\beta }$, and $n_{i}=\sum_{\sigma
}c_{i\sigma }^{\dag }c_{i\sigma }$ with $\langle n_{i}\rangle =1-x$ being
the number density operator. In addition to the Hamiltonian (\ref{H}) there
is the constraint of no double occupancy, which accounts for strong electron
correlations. The values of the parameters of the Hamiltonian (\ref{H}) for
LSCO are known from neutron scattering~\cite{Keimer92}, Raman spectroscopy~
\cite{tokura90}, and ab-initio calculations~\cite{andersen95} to be:%
\begin{eqnarray}
J &\approx &140\,\text{meV},\qquad t\approx 450\,\text{meV},  \label{values}
\\
t^{\prime } &\approx &70\,\text{meV},\qquad t^{\prime \prime }\approx 35\,%
\text{meV}.  \notag
\end{eqnarray}%
Note that the signs of the hopping terms in the electron Hamiltonian (\ref{H}%
) have been chosen in such a way that, for a $d_{x^{2}-y^{2}}$ orbital, the
hopping matrix elements are positive.

The dispersion of the hole dressed by magnetic quantum fluctuations has
minima at the nodal points $\mathbf{q}_{0}=(\pm \pi /2,\pm \pi /2)$ and is
practically isotropic in the vicinity of each~\cite{luscher06}: 
\begin{eqnarray}
\epsilon \left( \mathbf{q}\right) &\approx &\epsilon \left( \mathbf{q}%
_{0}\right) +\frac{1}{2}\beta (\mathbf{q}-\mathbf{q}_{0})^{2}  \label{eq} \\
\beta &\approx &2J\approx 260\,\text{meV}\ .  \notag
\end{eqnarray}%
We set the lattice spacing to unity, 3.81\thinspace \AA $\,\rightarrow $%
\thinspace 1. The effective mass corresponding to the quadratic dispersion (%
\ref{eq}) is approximately twice the electron mass.

\begin{figure}[h]
\centering
\par
\includegraphics[height=100pt, keepaspectratio=true]{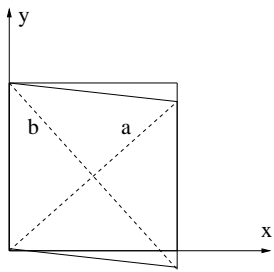}
\par
\caption{Orthorhombic deformation of the square lattice}
\label{Fig1}
\end{figure}

In the low-temperature orthorhombic (LTO) phase, the square Cu lattice is
slightly deformed as indicated in Fig.~\ref{Fig1}: The angles, $\pi /2\pm
\varphi ,$ between the edges are obtuse or acute, such that the $\left(11\right) $ 
and $\left( 1-1\right) $ diagonals, the orthorhombic translations,
have slightly different lengths$,$ respectively $a=\sqrt{2}\left( 1-\varphi
/2\right) $ and $b=\sqrt{2}\left( 1+\varphi /2\right) .$ On top of this
comes an alternating tilt of the oxygen octahedra around the $a$ axis by an
angle $\pm \theta ,$ whereby the oxygens in the layer buckle out of the Cu
plane forming a $\left[ -\pi /2,\pi /2\right] $-wave in the $b$-direction~%
\cite{Keimer92,lat}. The values of the orthorhombic and tilting
deformations,%
\begin{equation}
\varphi =0.009~\left( 0.5%
{{}^\circ}%
\right) \qquad \mathrm{and}\qquad \theta =0.05~\left( 3%
{{}^\circ}%
\right) ,  \label{def}
\end{equation}%
are so tiny that they hardly influence $J$, $t$ and $t^{\prime \prime },$
but they do make the diagonal hopping significantly different in the $a$ and 
$b$ directions, as we show below. Including $t_{a}^{\prime }\neq
t_{b}^{\prime }$ in the first, one-electron part of the Hamiltonian (\ref{H})
 yields the following 2D bandstructure:%
\begin{eqnarray}
\varepsilon \left( \mathbf{k}\right)  &=&-2t\left( \cos k_{x}+\cos
k_{y}\right) -2t^{\prime \prime }\left( \cos 2k_{x}+\cos 2k_{y}\right)  
\notag \\
&&+2t_{a}^{\prime }\cos \left( k_{x}+k_{y}\right) +2t_{b}^{\prime }\cos
\left( k_{x}-k_{y}\right) .  \label{H1}
\end{eqnarray}%
Note that $\varepsilon $ denotes electron energies and $\epsilon $ hole
energies. The splitting of the bare valleys is:%
\begin{eqnarray}
\varepsilon \left( -\pi /2,\pi /2\right) -\varepsilon \left( \pi /2,\pi
/2\right)  &\equiv &\varepsilon _{b}-\varepsilon _{a}  \label{valleysplit} \\
&=&4\left( t_{a}^{\prime }-t_{b}^{\prime }\right) \equiv 4\delta t^{\prime }.
\notag
\end{eqnarray}

\begin{figure}[h]
\centering
\par
\includegraphics[height=170pt, keepaspectratio=true]{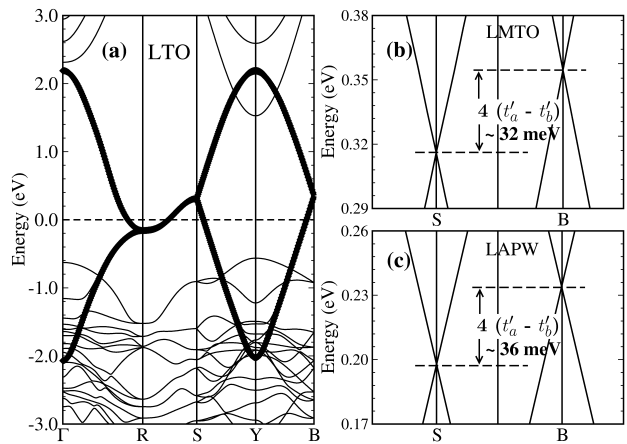}
\par
\caption{
LDA bandstructure of the LTO phase shown in the orthorhombic BZ
(see text). (a) NMTO downfolded Cu $d_{x^{2}-y^{2}}$ - like conduction bands
shown in thick lines\protect~\cite{nmto}. (b) Blow-up of conduction bands
near the nodal points S and B, and along the respective nodal lines, $a$ and 
$b.$ (c) same as (b), but calculated with the full-potential 
Linear Augmented Plane Wave Method (LAPW)
\protect~\cite{wien2k} instead of the 
Linear Muffin Tin Orbital Method (LMTO)\protect~\cite{tblmto}%
}
\label{bands}
\end{figure}

For comparison, we show in the left-hand side of Fig.\thinspace \ref{bands}
the bandstructure calculated from density-functional theory (LDA) along the
lines connecting the high-symmetry points $\Gamma $ $\left( 0,0,0\right) ,$
R $\left( \pi ,0,\pi /2\right) ,$ S $\left( \pi /2,\pi /2,\pi /2\right) ,$ Y 
$\left( 0,0,\pi /2\right) ,$ and B $\left( -\pi /2,\pi /2,\pi /2\right) $ of
the 3D \emph{orthorhombic} ($Bmab$) Brillouin zone (BZ). For simplicity, we
have given the coordinates in terms of the nearly \emph{tetragonal}
reciprocal Cu-lattice translations. $k_{z}=\pi /2$ is halfway to the
Brillouin-zone boundary where the influence of inter-layer hopping is
minimal. We see that no gapping is caused by the $\left[ -\pi /2,\pi /2%
\right] $ tilting wave. As a consequence, the LDA bandstructure may be
folded out to the nearly tetragonal BZ where expression (\ref{H1})
adequately represents the dispersion of the LDA conduction band, indicated
by heavy lining, near half-filling. On the righ-hand side of Fig.\thinspace %
\ref{bands} we show a blow-up of the orthorhombic LDA bands near the nodal
points, S $\left( \pi /2,\pi /2,\pi /2\right) $ and B $\left( \pi /2,-\pi
/2,\pi /2\right) ,$ along the respective nodal direction, $a$ and $b$.
Finally we see that two different computational techniques, LMTO and LAPW,
give essentially the same result, namely%
\begin{equation}
\delta t^{\prime }\approx 8\,\text{meV,\ }\;\mathrm{i.e.}\qquad \delta
t^{\prime }/t^{\prime }\approx 11\,\%.  \label{LDA}
\end{equation}%
This number is surprisingly large: Had the Cu $d_{x^{2}-y^{2}}$
conduction-band Wannier-like orbital been a simple, canonical orbital~\cite%
{Canonical78}, the hopping between two such orbitals would have decreased as
their distance to the power $-\left( l+l^{\prime }+1\right) $, which for two 
$d$ orbitals is $-5$, and, hence, $\delta t^{\prime }/t^{\prime }=5\varphi
=4.5\%.$ This, however, neglects that the hopping is almost exclusively via
the O$_{x}$ $p_{x}$ and O$_{y}\,p_{y}$ orbitals, as is amply demonstrated by
an LDA calculation in which we took the tilting wave to run along $a$
instead of along $b.$ The result was a four times reduction of $\delta
t^{\prime }$ compared to (\ref{LDA})!
 We therefore conclude that tilting is almost as important as
orthorhombicity for the hopping anisotropy.

To understand why, we turn to the simplest model showing an effect of
tilting: Emery's 3-band model~\cite{Em} which keeps merely the $t_{pd}$
hopping between Cu $d_{x^{2}-y^{2}}$ and its nearest O$_{x}$ $p_{x}$ and O$%
_{y}\,p_{y}$ orbitals, as well as the $t_{pp}$ hopping between the nearest O$%
_{x}$ $p_{x}$ and O$_{y}\,p_{y}$ neighbors. It is easy to see that in this
model,%
\begin{equation}
t^{\prime }=2\frac{t_{pd}t_{pp}t_{pd}}{\left( \varepsilon _{F}-\varepsilon
_{p}\right) ^{2}},\;\;\mathrm{so}\;\mathrm{that}\qquad \frac{\delta
t^{\prime }}{t^{\prime }}=\frac{\delta t_{pp}}{t_{pp}}  \label{tpp}
\end{equation}%
because $t_{pd}$ and the distance of the O $p$ level, $\varepsilon _{p},$
below the half-filling level, $\varepsilon _{F},$ are not influenced by the
deformations. In the deformed structure, each plane oxygen remains midway
between its two nearest Cu neighbors, but is slightly above or below the Cu
plane. Moreover, the O$\,p$ orbital is parallel to the Cu-Cu line, so that
the two O$\,p$ orbitals via which the $t_{a/b}^{\prime }$ hopping takes
place are at the acute/obtuse angle $\pi /2\mp \varphi .$ The main effect of
the deformation is, however, not this misalignment, but simply that the
distance between the oxygens at the acute angle is $a/2$ and that between
the oxygens at the oblate angle is $b/2/\cos \theta .$ With $l=l^{\prime }=1,
$ this gives: $\delta t^{\prime }/t^{\prime}=3\left( \varphi +\theta ^{2}\right)
=3\left( 0.90+0.25\right) \%\approx 3.5\%,$ which, due to the slower decay
of canonical $p$ orbitals, is even smaller than the previous $dd$ estimate.
Also the tilting contribution is too small. Inclusion of the orbital
misalignments in the canonical approximation~\cite{Canonical78} merely changes
the result to: $\delta t^{\prime }/t^{\prime}=\frac{7}{3}\varphi +5\theta ^{2}\approx
\left( 2.1+1.3\right) \%=3.4\%.$

Another simple model is the axial-orbital model~\cite{andersen95} which adds
to the three orbitals of the Emery model a Cu-centered, axial orbital, but
keeps only hops between nearest neighbors: from O\thinspace $p$ to Cu $%
d_{x^{2}-y^{2}}$ $\left( t_{pd}\right) $ and to the axial orbital $\left(
t_{sp}\right) .$ The axial orbital is a hybrid between Cu$\,4s,$ Cu $%
3d_{3z^{2}-1},$ apical O\thinspace $p_{z}$ and axial cation orbitals, and
its energy, $\varepsilon _{s}$ $\left( >\varepsilon _{F}\right) $ is \emph{%
the} material dependent electronic parameter in the LDA. Since $t_{pp}$
proceeds via the axial orbital in this model, and therefore equals $%
t_{4sp}^{2}/\left( \varepsilon _{s}-\varepsilon _{F}\right) ,$ it cannot
depend on $\varphi !$ For the same reason, the hopping from O$_{x}\,p_{x}$
to O$_{x}\,p_{x}$ over the distance 1 has the same value, $t_{pp},$ and this
is what leads to $t^{\prime \prime }=t^{\prime }/2$ when downfolding to the
1-band model~\cite{andersen95}.

However, recent first-principles 3-band Hamiltonians formed by numerical
downfolding of the LDA Hilbert space~\cite{nmto} do not support this; they
yield a $p_{x}$-$p_{x}$ hop, which can only be understood by the presence
of a large material-independent contribution, $-t_{4pp}^{2}/\left(
\varepsilon _{4p}-\varepsilon _{F}\right) ,$ from hopping via Cu 4$p_{x}$ 
\cite{H3}. To the $p_{x}$-$p_{y}$ hop, there can be no such contribution, $-$
\emph{unless} there is orthorhombic distortion. Also buckling-induced
anisotropy may be caused by hopping via Cu $4p_{z}$.

To see whether coupling via excited Cu $4p$ degrees of freedom can be the
reason for the surprisingly large anisotropy (\ref{LDA}) found by the LDA
calculations, we now add these degrees of freedom to the axial model and, as
usual, include hops only between nearest neighbors. For the directional
dependences we use the canonical approximation and find:%
\begin{eqnarray*}
\frac{\delta t^{\prime }}{t^{\prime }}=\frac{\delta t_{pp}}{t_{pp}}
&=&\left( 2\varphi +4.5\theta ^{2}\right) \left( \frac{t_{4pp}}{t_{sp}}%
\right) ^{2}\frac{\varepsilon _{s}-\varepsilon }{\varepsilon
_{4p}-\varepsilon } \\
&\approx &\left( 1.8+1.1\right) \%\times \left( \frac{2.6}{2.3}\right) ^{2}%
\frac{35}{11}=12\%.
\end{eqnarray*}%
The superb agreement with the LDA result (\ref{LDA}) may be fortuitous, but
also the relative contribution from orthorhombicity and tilt agrees. So we
believe that the surprisingly large hopping anisotropy to be explained by
this simple expression. It may be noted that going to materials with higher $%
T_{c\max },$ $\varepsilon _{s}$ decreases and with it the hopping anisotropy.
Finally it should be noted that for simplicity we have taken $\varepsilon
_{4z}=\varepsilon _{4p}\,\left( \equiv \varepsilon _{4x/y}\right) ,$
although $\varepsilon _{4z}$ is an axial orbital and, hence, material
dependent.

The $t^{\prime }$ describes hopping within the same magnetic sublattice.
Therefore, to account for many electron correlations in the anisotropy of the
single hole dispersion  one only needs to know the quasiparticle residue
$Z\left( \mathbf{q}\right) $.
This leads to anisotropic correction to the hole
dispersion~\cite{luscher06}%
\begin{equation}
\delta \epsilon \left( \mathbf{q}\right) =2Z\left( \mathbf{q}\right)
\,\delta t^{\prime }\sin q_{x}\,\sin q_{y}\ .  \label{de}
\end{equation}%
 The correction to the dispersion vanishes at
antinodal points $(0,\pi )$, $(\pi ,0)$ and it is maximum at nodal points, $%
(\pm \pi /2,\pm \pi /2)$, where $Z\approx 0.3$. The energy difference
between the nodal points is

\begin{equation}
\epsilon _{a}-\epsilon _{b}=4Z\delta t^{\prime }\sim 10\,\text{meV}\ .
\label{deab}
\end{equation}%
Thus, at low temperature all hydrogen-like bound states are formed from the $%
b$-valley holes, the $a$-valley is empty. We would like to stress that this
is true only in the insulating phase. In the superconducting phase which
arises after percolation of the bound states, $x>0.055$, both valleys are
populated.

The energy difference (\ref{deab}) is the single hole effect. It accounts
for spin quantum fluctuations, but it assumes usual Neel order. There is
another collective contribution to the energy difference $\epsilon
_{a}-\epsilon _{b}$. The collective contribution is due to the spiral spin
ordering established at $0.02\leq x\leq 0.055$. The collective contribution
is considered below in the ARPES section. Certainly, in the end the
collective contribution is also driven by the asymmetry (\ref{deab}) because
the asymmetry's causes depopulation of the $a$-valley.

\section{Implications for optical transitions $p^5d^9 \to p^6d^8$}
\begin{figure}[th]
\includegraphics[width=0.6\textwidth,clip]{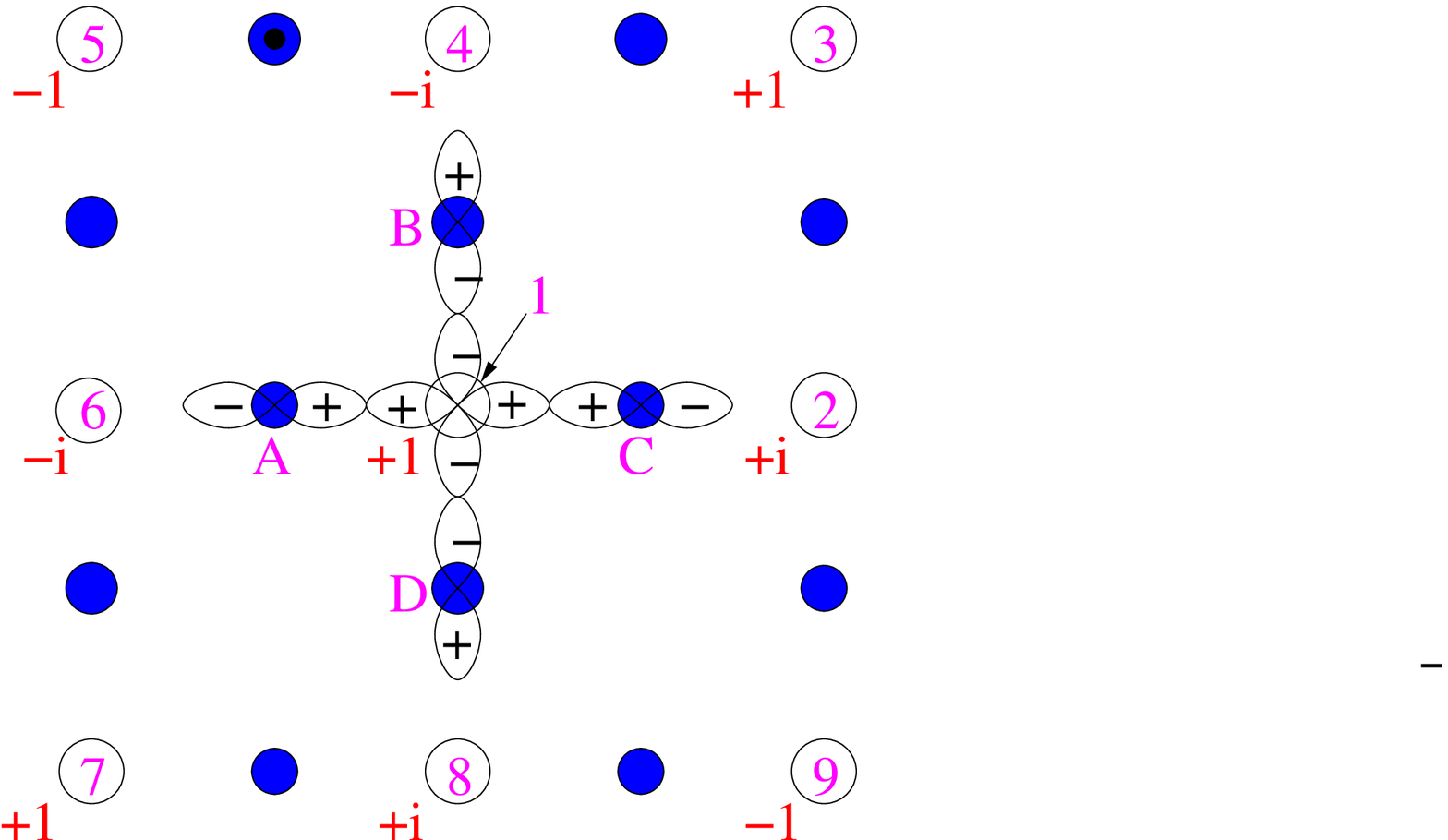}
\caption{(Color online) CuO$_{2}$ plane. Open circles denote Cu ions and
blue circles denote Oxygen ions. Cu ions in the cluster are enumerated by
magenta numbers 1-9 and Oxygen ions are enumerated by
four magenta letters  A,B,C,D. The phase factor, 
$\pm 1,\pm i$, is shown in red near the corresponding Cu ion.}
\label{figA}
\end{figure}

We consider the optical transitions involving removing a $d$ electron from a 
central placket of Fig.~\ref{figA} resulting in a $d^8$-state and transferring
it to a linear combination of Zhang Rice (ZR) singlet states on neighboring 
plackets. Note that the transition staying within the same plaque is not allowed
optically. This particular transition is of course only allowed if there is
a hole in the ZR state on neighboring plackets and therefore is only present in a 
hole doped material. In the undoped material the corresponding transition 
would have to be to neighboring d states resulting in a linear combination of 
$d^{10}$ configurations there. This transition would be at an energy of about 
the charge transfer gap higher than the transitions to ZR singlet states and 
is easily distinguishable from those to the ZR states. These transitions 
involve the full $d^8$ multiplet structure and their energies can be obtained 
from the calculations described in Ref.~\cite{ES}
as well as from early resonant photoemission experiments which locate the 
$d^8$ multiplets in these materials.~\cite{TCC}
Looking at Fig. 1 of Ref.~\cite{ES} we see that the $d^8$ states occur in the 
energy range between 8 and 15 eV below the ZR singlet state. The large energy 
spread is due to the multiplet structure in the $d^8$ configurations. 

We now describe why the polarization dependence of  these transitions will be 
very sensitive to where in momentum space the ZR states are situated and 
therefore the polarization dependence represents a test of the hypothesis 
described above. 
For this discussion the orthorhombic distortion is no longer important except 
for the proposal that it causes the doped hole states to be concentrated at 
the b minimum with a wave function given by 
\begin{equation}
\psi _{b}(\mathbf{r})=\chi (\mathbf{r})\exp \left\{ i\frac{\pi }{2}x-i\frac{%
\pi }{2}y\right\} \ ,
\end{equation}%
where $\chi (\mathbf{r})\propto e^{-\kappa r}$ is a relatively smooth wave
function of the bound state, $\kappa \approx 0.4$. The smooth wave function
is not important for optical absorption. Only the fast phase factor $\exp
\left\{ i\frac{\pi }{2}x-i\frac{\pi }{2}y\right\} $ is important. For a
bound state based on a hole from the $a$-minimum the function $\chi (\mathbf{%
r})$ is the same while the phase factor $\exp \left\{ i\frac{\pi }{2}x+i%
\frac{\pi }{2}y\right\} $ is different.

To describe the above transition we clearly need to go beyond a $t-J$ or 
Hubbard model taking into account explicitly the charge transfer nature of 
the gap as in the Zaanen-Sawatzky-Allen classification scheme~\cite{ZSA}.
In Fig.~\ref{figA} we display the CuO$_2$ plane structure and the orbitals 
considered. The hole phase factor 
$\exp \left\{ i\frac{\pi }{2}x-i\frac{\pi }{2}y\right\}
=\pm 1,\pm i$ is shown in red near the corresponding Cu ion. The transition to a 
zero momentum ZR state is parity forbidden. 
However the hole resides in a ZR state with nonzero momentum where
transition is optically allowed due to interference of ZR states centered on 
different Cu sites.
 Let us consider the case when the final hole resides at
the first Cu site, $|final\rangle =|{\bar{d}}_{1}\rangle $.
Here for short notations we denote $d^8$ state as ${\bar d}$.
 The transition ${\bar{p}}\rightarrow {\bar{d}}_{1}$ can occur when the initial 
p-hole resides on Oxygens A,B,C,D and comes from the ZR states centered at Cu sites
2,4,6,8. Therefore the transition amplitude is
\begin{equation}
A=\langle {\bar{d}}_{1}|E_{x}x+E_{y}y|-i{\bar{p}}_{A}+i{\bar{p}}_{C}-i{\bar{p%
}}_{B}+i{\bar{p}}_{D}\rangle \ ,
\label{A}
\end{equation}
where ${\vec{E}}$ is the electric field of the photon. Let us denote by 
{\cal D} the dipole matrix element 
\begin{equation}
\langle {\bar{d}}_{1}|x|{\bar{p}}_{C}\rangle =\mathcal{D}=\int \psi
_{d_{1}}x\psi _{p_{C}}dV\ .
\end{equation}%
The symmetry relations between nonzero matrix elements follow from Fig.~\ref%
{figA} 
\begin{eqnarray}
&&\langle {\bar{d}}_{1}|x|{\bar{p}}_{A}\rangle =-\mathcal{D}\ ,  \notag \\
&&\langle {\bar{d}}_{1}|y|{\bar{p}}_{B}\rangle =\mathcal{D}\ ,  \notag \\
&&\langle {\bar{d}}_{1}|y|{\bar{p}}_{D}\rangle =-\mathcal{D}\ .
\end{eqnarray}%
Hence, we find from (\ref{A})%
\begin{equation}
A=2i(E_{x}-E_{y})\mathcal{D}\ .
\end{equation}%
Thus we conclude that there is an absorption if the wave is polarized along
the orthorhombic $b$-axis, ${\vec{E}}\propto (1,-1)$ and there is no
absorption if the wave is polarized along the orthorhombic $a$-axis, ${\vec{E%
}}\propto (1,1)$. Basically this is the only possible correlation one can
write kinematically, $I\propto ({\vec{E}}\cdot {\vec{k}})^{2}$, where ${\vec{%
k}}$ is momentum of the hole. So, the answer is obvious even without a
calculation. The optical absorption we have discussed is proportional to
doping $x$. We stress that the prediction for the low temperature absorption
asymmetry is $\sigma_b \gg \sigma_a$ and this is
equally applicable to the Neel and \textquotedblleft 
spin-glass\textquotedblright\ phases of LSCO. 
The optical asymmetry is different
from that in the dc conductivity where both experimentally~\cite{ando02} and
theoretically~\cite{kotov05} the asymmetry at low temperature is
not that large, $\sim $50\%, and of the opposite sign, 
$\sigma_b < \sigma_a$.

\section{Implications for ARPES}

We consider photoemission from $a$ and $b$ nodal points. There is a
difference in energy that is given by Eq. (\ref{deab}). This is already an
interesting effect. However, there is a much bigger effect that is due to
the spin spiral in the \textquotedblleft spin-glass\textquotedblright\
phase. The pitch of the spiral directed along the $b$-axis is~\cite%
{sushkov05,luscher07} 
\begin{equation}
\mathbf{Q}=\frac{gx}{\rho _{s}}(1,-1)\ ,  \label{p}
\end{equation}%
where $\rho _{s}\approx 0.18J$ is spin stiffness, $g\approx Zt\approx 0.7J$
is the hole-spin-wave coupling constant, and $x$ is doping. Eq. (\ref{p})
agrees very well with neutron scattering data. A hole with momentum $\mathbf{%
q}$ interacts with the spiral. The interaction splits the hole dispersion in
two branches with the following energy shift~\cite{sushkov05,luscher07} 
\begin{equation}
\Delta \epsilon \left( \mathbf{q}\right) =\pm g|Q_{x}\sin q_{x}+Q_{y}\sin
q_{y}|\ .  \label{dde}
\end{equation}%
Thus the hole dispersion near the \textquotedblleft $a$%
\textquotedblright\ nodal point, $\mathbf{q}\approx (\pi /2,\pi /2)$, is
practically not influenced by the spiral, $\Delta \epsilon _{a}=0$. On the
other hand the hole dispersion near the \textquotedblleft $b$%
\textquotedblright\ nodal point, $\mathbf{q}\approx (\pi /2,-\pi /2)$, is
changed as 
\begin{equation}
\Delta \epsilon _{b}=\pm \frac{2g^{2}x}{\rho _{s}}\ .  \label{ddea}
\end{equation}%
Accounting this correction together with (\ref{deab}) we find%
\begin{equation}
\epsilon _{b}-\epsilon _{a}=-4Z\delta t^{\prime }\pm \frac{2g^{2}x}{\rho _{s}}%
\sim -10\pm 770x\,\ \ \text{meV}\ .  \label{deab1}
\end{equation}%
Note that the second (collective) contribution of this difference scales
linearly with doping and it is pretty large, it is $\sim $30meV at $x=0.04$.

Thus the present picture predicts that the lowest branch of the dispersion
with energy approximately equal to chemical potential is at the nodal $b$%
-point. The dispersion at the nodal $a$-point is by $\sim $40meV higher (we
present the estimate for $x=0.04$). Finally, there is another branch of the
dispersion at the nodal $b$-point that is by $\sim $60meV above the chemical
potential, this branch can be pretty broad.
 The present consideration is applicable to the doping interval
$0.02\leq x\leq 0.055$ where the disordered spiral is established.
In the Neel phase, $%
x\leq 0.02$, the collective contribution in (\ref{deab1}) is suppressed.

\section{Conclusions}
We have demonstrated that the 0.9\% orthorhombic distortion of the Cu
lattice causes 7\% difference in diagonal hopping matrix elements, and that
the 3$%
{{}^\circ}%
$ tilting of the oxygen octahedra causes an additional 4\% difference. This
11\% difference together with the effect of strong magnetic fluctuations
(small hole pockets) and together with localization of holes due to Coulomb
trapping by Sr ions leads to depopulation of the $(\pi /2,\pi /2)$ pocket in
the insulating phase, $x\leq 0.055$. As a result the optical transition $%
p^{5}d^{9}\rightarrow p^{6}d^{8}$ is allowed only if the electric field is
polarized along the orthorhombic $b$-axis.

Another prediction that is related to the spin spiral structure is the
asymmetry of ARPES spectra: the hole dispersion at the $b$-nodal point is
close to the chemical potential, while at the $a$-nodal point it is by 25-50
meV higher depending on doping.

\end{document}